\documentclass[fleqn,twoside]{article}

\usepackage{espcrc2}
\usepackage{graphicx}
\usepackage{axodraw}
\usepackage{amssymb}

\newcommand{\be}{\begin{equation}}
\newcommand{\bea}{\begin{eqnarray}}
\newcommand{\ee}{\end{equation}}
\newcommand{\eea}{\end{eqnarray}}
\newcommand{\bpi}{\begin{picture}}
\newcommand{\bce}{\begin{center}}
\newcommand{\epi}{\end{picture}}
\newcommand{\ece}{\end{center}}

\newcommand{\de}{d^{-1}}

\def\g{{\rm I}\hspace{-0.07cm}\Gamma}

\def\r#1{(\ref{#1})}

\title{The QCD Effective Charge to All Orders} 

\author{D. Binosi\address[UV]{Departamento de F\'\i sica
Te\'orica and IFIC, Centro Mixto,  
Universidad de Valencia-CSIC,
E-46100, Burjassot, Valencia, Spain} and J. Papavassiliou\addressmark[UV]}

\begin{document}

\begin{abstract}
Using
the pinch technique, we show how to construct the QCD effective charge
to all orders. 
\end{abstract}

\maketitle

The extension  of the  concept of  the renormalization-group-invariant  
and process-independent effective charge  from QED to
QCD  is of special  interest for several reasons.   
First, the
existence of a QED-like effective  charge, and the ability to identify
directly   and   unambiguously    the   infinite   subset   of   gluon
self-energy-like  radiative  corrections,   is  assumed  in  renormalon
analysis of the behavior of the perturbative QCD series at high
orders, and in particular, in the study of nonperturbative effects 
near the QCD  mass scale \cite{Beneke:1998ui}. 
Moreover, such an effective charge could serve as the natural scheme for
defining  the  coupling in  the  recently  proposed ``event  amplitude
generator'' based on the light-cone formulation of QCD
\cite{Brodsky:2001wx}.
Furthermore, 
when studying the interface  between perturbative and non perturbative
effects in  QCD, one  finds it advantageous  to use  this familiar
QED concept, in conjunction with dispersive
techniques and analicity properties of the $S$-matrix 
\cite{Dokshitzer:1995qm}.

In QED, the effective charge construction proceeds as follows 
\cite{Itzykson:rh}: One begins by considering  
the unrenormalized photon self-energy   $\Pi^{\rm o}_{\alpha\beta}(q) =  q^2
P_{\alpha\beta}(q)\Pi^{\rm o}(q^2)$,  where
$P_{\alpha\beta}(q)=g_{\alpha\beta}-q_\alpha q_\beta/q^2$ and  
$\Pi^{\rm o} (q^2)$ is  a
gauge-independent function to all  orders in perturbation theory.  After
Dyson   summation,  we obtain   the (process-independent)  dressed
photon propagator between conserved external currents
$\Delta^{\rm o}_{\alpha\beta}(q)\ = 
(g_{\alpha\beta}/q^2)\Delta^{\rm o}(q^2)$, with 
$\Delta^{\rm o}(q^2)= -i[1+i\Pi^{\rm o}(q^2)]^{-1}$.
The renormalization procedure introduces 
the standard relations  between  renormalized    and unrenormalized
parameters:  
$e = Z_{e}^{-1} e^{\rm o} = Z_f Z_A^{1/2} Z_1^{-1} e^{\rm o}$
and 
$1+i\Pi (q^2)= Z_A [1+i\Pi ^{\rm o} (q^2)]$, 
where $Z_A$ ($Z_f$) is the wave-function renormalization 
constants of the
photon (fermion),  $Z_1$ the vertex renormalization, 
and $Z_{e}$ is the charge renormalization constant. 
The  Abelian gauge
symmetry  of the   theory   gives  rise   to  the    fundamental 
Ward identity (WI) 
$q^{\alpha}\Gamma^{\rm o}_{\alpha}(p,p+q)= S_{\rm o}^{-1}(p+q)-S_{\rm
o}^{-1}(p)$, where 
$\Gamma^{\rm o}_{\alpha}$ and  $S_{\rm o} (k)$ are the  unrenormalized
all orders    
photon-electron vertex and electron propagator, respectively.
The requirement  that     the   renormalized vertex       $\Gamma_{\alpha} 
=
Z_{1}\Gamma^{\rm o}_{\alpha}$   and  the  renormalized  self-energy   $S  =
Z_{f}^{-1} S_{\rm o}$ satisfy the same identity, implies 
$Z_{1}=Z_{f}$, from which immediately follows that 
$Z_{e}\ =\ Z_{A}^{-1/2}$. Given these relations between 
the renormalization constants, and after pulling out the trivial
factor $g_{\alpha\beta}/q^2$, 
we can form the renormalization-group-invariant combination, known as
the effective charge, 
\be
\alpha(q^2)=
\frac{(e^{\rm o})^2}{4\pi}\Delta^{\rm o}(q^2)=
\frac{e^2}{4\pi} \Delta(q^2).  
\label{alphaqed}
\ee

In QCD, 
the crucial equality $Z_1=Z_f$ does not
hold, because the WIs are replaced by 
the more complicated Slavnov-Taylor identities (STIs),  
involving ghost Green's functions.  
Furthermore, the gluon
vacuum  polarization 
depends on  the gauge-fixing parameter, already  at the one-loop
order. These facts complicate the QCD definition of a universal
(process-independent) effective charge \cite{Watson:1996fg} 
(process-dependent effective
charges have been considered in \cite{Grunberg:1992mp}). However, the
theoretical framework of the 
pinch technique (PT) \cite{Cornwall:1982zr} made this 
definition possible. 

The PT is a
well defined algorithm which  
reorganizes     systematically    a given   physical   amplitude  into
sub-amplitudes,   which have   the   same   kinematic  properties   as
conventional $n$-point functions, (propagators, vertices, boxes)
but, in addition, are endowed with desirable physical properties.
Most importantly, they 
are independent of  the  gauge-fixing
parameter and satisfy naive,
(ghost-free) tree-level  WIs instead    of   the usual
STIs.
The basic observation, which essentially defines the PT, is that there
exists   a   fundamental  cancellation,   driven   by  the   underlying
Becchi-Rouet-Stora-Tyutin symmetry, which
takes  place  between  sets   of  diagrams  with  different  kinematic
properties,  such  as self-energy,  vertex,  and  box diagrams.   This
cancellations  are  activated  when longitudinal  momenta  circulating
inside vertex and box diagrams, generate (by ``pinching'' out internal
fermion lines) propagator-like terms; the latter are combined with the
conventional self-energy graphs in order  to give rise to an effective
gluon  self-energy.
Recently, the PT has been generalized
to all orders in perturbation theory
through the collective treatment  of entire sets
of  diagrams \cite{Binosi:2002ft}. 

In this talk we will outline the all-order construction of the 
QCD effective charge. 
This is best done in the context of the {\it intrinsic} 
PT procedure.
The basic idea is
that the ``pinch graphs'', 
which are essential in  canceling the gauge dependences
of ordinary diagrams, are always missing one or more (tree-level)
propagators corresponding 
to the external legs of the improper Green's function in question. It then
follows that the gauge-dependent parts of such ordinary diagrams must also be
missing one or more external propagators. Thus the goal
of the intrinsic PT construction
is to isolate systematically the parts of the diagrams that are 
proportional to the inverse propagators of the external legs and simply discard
them. In fact, this is absolutely equivalent
to the usual ($S$-matrix) PT
construction: discarding these terms is justified because 
we know that inside an $S$-matrix element
they will eventually cancel (to all orders) 
against similar 
pieces stemming from the vertices \cite{Binosi:2002ft}.
The important point is that these inverse propagators  arise from the
STI satisfied by the (all-order)
three-gluon vertex appearing inside appropriate sets of
diagrams, when it is contracted by longitudinal momenta. Denoting by
$\g_{A_\alpha A_\mu A_\nu}(q,k_1,k_2)$, this STI reads \cite{Ball:ax} 
\bea
& &\!\!\!\!\!\!\!\!\!\!\!
k_1^\mu\g_{A_\alpha A_\mu A_\nu}(q,k_1,k_2)=  \nonumber \\
& &\!\!\!\!\!\!\!\!\!\!\! 
\left[i\Delta^{(-1)\,\rho}_{\nu}(k_2)+k_{2}^{\rho}k_{2\nu}\right]
\left[k_1^2D(k_1)\right]H_{\rho\alpha}(k_2,q) \nonumber \\
& &\!\!\!\!\!\!\!\!\!\!\!- 
\left[i\Delta^{(-1)\,\rho}_{\alpha}(q)+q^\rho q_\alpha\right]
\left[k_1^2D(k_1)\right]H_{\rho\nu}(q,k_2),
\label{STIbc}
\eea  
where $H$ represents the ghost Green's function appearing in the
conventional formalism; at tree level
$H_{\alpha\beta}^{[0]}(k_1,k_2)=-igg_{\alpha\beta}$.

Without loss of generality one can choose the
Feynman gauge; then, the only source of longitudinal momenta will be
the term $\Gamma^{\rm P}$ appearing in the decomposition  
$\g_{A_\alpha A_\mu
A_\nu}^{[0]}=\Gamma^{{\rm
F}}_{\alpha\mu\nu}+ \Gamma^{{\rm
P}}_{\alpha\mu\nu}$ with~\cite{Cornwall:1982zr}     
\bea
\Gamma^{{\rm
F}}_{\alpha\mu\nu}&=&(k_1-k_2)_\alpha  g_{\mu\nu}+2q_\nu
g_{\alpha\mu}-2q_\mu g_{\alpha\nu},\nonumber    \\    
\Gamma^{{\rm
P}}_{\alpha\mu\nu}&=&k_{2\nu}g_{\alpha\mu}-
k_{1\mu}g_{\alpha\nu}.
\label{decomp}
\eea

On the other hand, with the help of the Batalin-Vilkovisky (BV) formalism
\cite{Batalin:pb}, formulated in the Feynman gauge 
of the background field method (BFM) \cite{Abbott:1981ke},
 one can relate the BFM gluon two-point function
$\widetilde\g_{\widetilde A_\alpha \widetilde A_\beta}$ 
($\widetilde A$ denotes the background gluon)
with the
conventional one 
$\g_{A_\alpha A_\beta}$ through the background-quantum identity (BQI)
\bea
\widetilde\g_{\widetilde A_\alpha \widetilde A_\beta}(q)\!&=&\!\g_{A_\alpha
A_\beta}(q) 
+2\g_{\Omega_\alpha A^*_\rho}(q)\g_{A^\rho A_\beta}(q)\nonumber \\
&+&\!\g_{\Omega_\alpha A^*_\rho}(q)\g_{A^\rho A^\sigma}(q)\g_{\Omega_\beta
A^*_\sigma}(q), 
\label{BQI}
\eea
where $\g_{A_\alpha A_\beta}^{[0]}(q)=
-iq^2P_{\alpha\beta}(q)$, 
$\g_{A_\alpha A_\beta}^{[n]}(q)=\Pi^{[n]}_{\alpha\beta}(q^2)$,  
and $\g_{\Omega A^*}$ represents an auxiliary (unphysical) two-point
function connecting a background source $\Omega$ with a gluon
anti-field $A^*$ (see \cite{Binosi:2002ez} and references therein  
for details). The crucial observation made in \cite{Binosi:2002ez} has
then been that 
the STI of Eq.\r{STIbc} and  the BQI of Eq.\r{BQI} are related,
since the different auxiliary Green's function appearing in them
are connected by a Schwinger-Dyson type of relation, which reads
\bea
i\g_{\Omega_\alpha A^*_\beta}(q)&=&C_A 
\int
H^{[0]}_{\alpha\rho}(q,-k-q)D(k)\times\nonumber \\
&\times&\Delta^{\rho\sigma}(k+q)H_{\beta\sigma}(-q,k+q),
\label{gpert2}
\eea
where $C_A$ denotes the Casimir eigenvalue of the adjoint
representation, {\it i.e.}, $C_A=N$ for $SU(N)$, and 
we have defined the integral measure 
\mbox{$\int\equiv\mu^{2\varepsilon}\int\!\frac{d^dk}{(2\pi)^d}$}, with
$d=4-2\varepsilon$ and $\mu$ the 't~Hooft mass. 

After these observations, we proceed to the 
construction of the PT gauge independent
two-point function $\widehat \g_{A_\alpha A_\beta}$; as we will
show, it
coincides with the corresponding BFM
two-point function $\widetilde 
\g_{\widetilde A_\alpha \widetilde A_\beta}$. 
The one particle irreducible (1PI) 
Feynman  diagrams contributing  to the  conventional 
gluon self-energy in the $R_\xi$  gauges can be always 
separated into three distinct
sets:  ({\it  i})   the  set  of  diagrams  that   have  two  external
(tree-level)   three-gluon   vertices,  and   thus   can  be   written
schematically  (suppressing  Lorentz  indices) as  $\Gamma^{[0]}[{\cal
K}_2]\Gamma^{[0]}$, where ${\cal K}_2$ is some kernel; ({\it ii}) the set
of diagrams  with only  one external (tree-level)  three-gluon vertex,
and  thus  can  be  written  as $\Gamma^{[0]}[{\cal  K}_1]$  or  $[{\cal
K}_1]\Gamma^{[0]}$;  ({\it iii}) all  remaining diagrams,  containing no
external three-gluon vertices. Then, if we carry out  the
decomposition 
$\Gamma^{[0]}\Gamma^{[0]}=\Gamma^{\rm F}
\Gamma^{\rm F}+\Gamma^{\rm P}\Gamma^{[0]}+\Gamma^{[0]}\Gamma^{\rm P} 
-\Gamma^{\rm P}\Gamma^{\rm P}$ 
to the pair of external  vertices appearing in
the diagrams of the set ({\it i}), and the decomposition of
Eq.(\ref{decomp}) 
to the
external vertex appearing in the diagrams of the set ({\it ii}), 
after a judicious
rearrangement of the kernels ${\cal K}_2$ and ${\cal K}_1$ (together with their
statistical factors), relabeling of internal 
momenta,  and
taking into account the transversality of the gluon self-energy, 
the pinching contribution coming from the 1PI diagrams will read
\bea
\left\{\g_{A_\alpha A_\beta}\right\}^{\rm
P}\!\!\!&=&\!\!\!2iC_A\int\!\frac1{k^2}\Gamma^{\rm
P}_{\alpha\mu\nu}(q,k,-k-q)\times \nonumber \\
\!\!\!&\times&\!\!\!\Delta^\nu_\sigma(k+q) \g_{A_\beta A^\mu
A^\sigma}(q,k,-k-q). \nonumber \\ 
\label{1PIPT}
\eea

Thus, the longitudinal terms $k_{\mu}$ and $(k+q)_{\nu}$
stemming from  $\Gamma^{{\rm P}}_{\alpha \mu \nu}$   
will trigger the STI of Eq.(\ref{STIbc}). 
Therefore,  the all order generalization of the
intrinsic PT will amount to  isolating from 
Eq.\r{1PIPT} the terms of the STI of Eq.(\ref{STIbc}) 
that are proportional to 
$[\Delta^{(-1)\,\rho}_{\alpha}(q)]$; we will 
denote such contributions by  $\Pi_{\alpha\beta}^{{\rm
IP}}(q)$. Thus the 1PI diagrams contributing to the
gluon self-energy can be cast in the form 
$\g_{A_\alpha A_\beta}(q)=G_{A_\alpha A_\beta}(q)+
\Pi_{\alpha\beta}^{{\rm IP}}(q)$.

Notice however that   
the one particle reducible (1PR) 
set ${\mathbb S}$ containing  diagrams constructed from
strings of lower order self-energy graphs 
must also be rearranged following the intrinsic PT procedure, and be
converted into the equivalent set $\widehat{\mathbb S}$ containing 
strings involving PT self-energies. 
This treatment of the 1PR strings will
give rise, in addition to the PT strings, 
to ({\it a}) a set of contributions which are proportional to the
inverse tree-level propagator of the external legs $\de(q)$ (with
$d(q)=-i/q^2$), and ({\it b})
a set of contributions which is {\it
effectively} 1PI, and therefore  also belongs to the definition of the 1PI
PT gluon self-energy; we will denote these two sets of
contributions 
collectively by $S^{{\rm IP}}_{\alpha\beta}(q)$. Thus the sum of 
the 1PI and 1PR contributions to the conventional gluon  
self-energy can be cast in the form  
\bea
\g_{A_\alpha A_\beta}(q)+{\mathbb S}_{\alpha\beta}(q)
&=&G_{A_\alpha A_\beta}(q)+\widehat{\mathbb
S}_{\alpha\beta}(q)\nonumber \\
&+&
\Pi_{\alpha\beta}^{{\rm IP}}(q)+S^{{\rm
IP}}_{\alpha\beta}(q).
\label{ip2}
\eea

By definition of the intrinsic PT procedure, we will now discard from the above
expression all the terms which are proportional to the inverse
propagator of the external legs, 
thus defining the quantity $R^{{\rm
IP}}_{\alpha\beta}(q)=\Pi'^{\,{\rm IP}}_{\alpha\beta}(q)+S'^{\,{\rm
IP}}_{\alpha\beta}(q)$, 
where the primed functions are defined starting from the unprimed ones
appearing in Eq.\r{ip2} by discarding the aforementioned terms. 
Thus, finally,
the all orders intrinsic PT gluon
self-energy ${\widehat\g}_{A_\alpha
A_\beta}(q)$, will be defined as
\bea
{\widehat\g}_{A_\alpha A_\beta}(q)&=& G_{A_\alpha
A_\beta}(q)+R^{{\rm IP}}_{\alpha\beta}(q)\nonumber \\
&=&\g_{A_\alpha A_\beta}(q) - \Pi_{\alpha\beta}^{{\rm IP}}(q) +
R^{{\rm IP}}_{\alpha\beta}(q).\nonumber 
\eea

We next proceed to the construction of the quantities
$\Pi_{\alpha\beta}^{{\rm IP}}(q)$ and $R^{{\rm IP}}_{\alpha\beta}(q)$
discussed above, starting from the first one.
Using the definition of $\Gamma^{\rm P}$ of Eq.\r{decomp}, together with the
STI \r{STIbc} (keeping only pinching terms) and the tree-level
value of the $H$ Green's function, from Eq.\r{1PIPT} we get 
\bea
\Pi^{\rm
IP}_{\alpha\beta}(q)\!\!\!&=&\!\!\!2iC_A\int\!H^{[0]}_{\alpha\nu}(q,-k-q)
D(k)\times \nonumber \\
&\times&\!\!\!
\Delta^{\nu\sigma}(k+q)H_{\rho\sigma}(-q,k+q)\g_{A^\rho
A_\beta}(q), \nonumber 
\eea
which, using Eq.\r{gpert2}, implies 
$\Pi^{\rm
IP}_{\alpha\beta}(q)=-2\g_{\Omega_\alpha A^*_\rho}(q)\g_{A^\rho
A_\beta}(q)$.

From the 1PR set of diagrams ${\mathbb S}_{\alpha\beta}$ instead, 
we need to
identify the subset of contributions $S^{\rm{IP}}_{\alpha\beta}$
which is effectively 1PI. In what follows  
we will suppress Lorentz and momentum indices. Now, it can be shown
that the only elements of the 1PR
set $\mathbb S$ that can contribute to $S^{\rm{IP}}$ are the strings that
contains at most three self-energy insertions, {\it i.e.}, the subsets
${\mathbb S}_2=\g_{AA}\,d\,\g_{AA}$ and 
${\mathbb S}_3=\g_{AA}\,d\,\g_{AA}\,d\,\g_{AA}$
\cite{Papavassiliou:1995fq}. 
The reason for this is that the terms that one needs to add to a string of
order $n$ containing more than three self-energy insertions to
convert it into a PT string, will be exactly 
canceled by the conversion into PT strings 
of other strings of the same order, but containing a
different number of insertions.
The only time that this will not happen is when 
the string contains two or three self-energy insertions. In
this case we will get
\bea
{\mathbb S}^{[n]}_2+{\mathbb S}^{[n]}_3
&\to&\widehat{\mathbb S}^{[n]}_2+\widehat{\mathbb S}^{[n]}_3+
2\sum_{m=1}^{n-1}\g_{\Omega A^*}^{[n-m]}\g_{AA}^{[m]}\nonumber \\
&+&
\sum_{m=1}^{n-1}\sum_{\ell=0}^{m-1}
\g_{\Omega A^*}^{[n-m]}\g_{AA}^{[\ell]}\g_{\Omega
A^*}^{[m-\ell]}\nonumber \\
&=& \widehat{\mathbb S}^{[n]}_2+\widehat{\mathbb S}^{[n]}_3+
S'^{\,{\rm IP}\,[n]}.
\eea
On the other hand, we have that 
$\Pi'^{\,{\rm
IP}\,[n]}=-2\sum_{m=1}^{n-1}
\g_{\Omega A^*}^{[n-m]}\g^{[m]}_{AA}$, 
so that putting back Lorentz and momentum indices, 
we get the all order result $R^{{\rm IP}}_{\alpha\beta}(q)=
\g_{\Omega_\alpha A^*_\mu}(q)\g_{A^\mu A^\nu}(q)
\g_{\Omega_\beta A^*_\nu}(q)$.
Thus, making use of the BQI of Eq.\r{BQI}, we have the identity
\bea
{\widehat\g}_{A_\alpha A_\beta}(q)&=&
\g_{A_\alpha A_\beta}(q) - \Pi_{\alpha\beta}^{{\rm IP}}(q) +
R^{{\rm IP}}_{\alpha\beta}(q)\nonumber \\
&=&\widetilde \g_{\widetilde A_\alpha \widetilde A_\beta}(q).
\eea

As a result we can define the unrenormalized PT gluon self-energy as
$\g_{A_\alpha A_\beta}=q^2P_{\alpha\beta}(q)\widehat\Pi^{\rm o}(q^2)$,
where we have shown that $\widehat\Pi^{\rm o}(q^2)$ is gauge independent to
all orders in perturbation theory. Thus, after Dyson summation, 
we obtain the dressed PT gluon propagator
between external conserved currents $\widehat\Delta_{\alpha\beta}^{\rm
o}(q)=(g_{\alpha\beta}/q^2)\widehat\Delta^{\rm o}(q^2)$, where 
$\widehat\Delta^{\rm o}(q^2)=-i[1+i\widehat\Pi^{\rm o}(q^2)]^{-1}$. On the
other hand, as it has been shown in \cite{Binosi:2002ft}, the bare PT
gluon-quark vertex $\widehat\Gamma^{{\rm o},\,a}_\alpha(q)$ coincides 
with the corresponding BFM Feynman gauge vertex $\widetilde\Gamma^{{\rm
o},\,a}_\alpha(q)$ to all orders in perturbation theory, ensuring the
validity of the 
equality $Z_1=Z_f$ also in the QCD case. Thus, 
in complete analogy to QED, we can form the
renormalization-group-invariant combination
\be
\alpha_s(q^2)=\frac{(g^{\rm o})^2}{4\pi}\widehat\Delta^{\rm o}(q^2)=
\frac{g^2}{4\pi}\widehat\Delta(q^2),
\ee
where $g$ is the QCD coupling constant.

In conclusion, we have presented the all-order construction of the 
effective charge of QCD, by means of the (intrinsic) PT
algorithm. 
This construction allows for the explicit identification of the
conformally (in)variant subsets of QCD graphs \cite{Brodsky:1982gc},
and lends itself as the natural candidate for 
studying issues such as dynamical gluon mass generation in a manifestly 
gauge-invariant way \cite{Cornwall:1982zr}.

{\bf Acknowledgments:}
This work has been supported by the CICYT Grants
AEN-99/0692 and BFM2001-0262. ~J.P. thanks the organizers of QCD~02
for their hospitality and for 
providing a very pleasant and stimulating atmosphere.

\end{document}